\documentclass{elsart5p}

\usepackage[dvips]{graphicx}
\usepackage{amssymb}
\usepackage{amsmath}
\usepackage{blackdvi}

\def \Kla#1{\left( #1 \right)}

\def \Klc#1{\left\{ #1 \right\}}

\def \Unit#1{\thinspace\hbox{\rm #1}}

\def \eH#1{{\rm e}^{#1}}
\def \Np{{N^+}}
\def \Nm{{N^-}}

\begin{document}
\setlength{\pdfpagewidth}{210 mm}
\setlength{\pdfpageheight}{297 mm}

\begin{frontmatter}



\title{Constraints on spin-dependent short-range interactions using
  gravitational quantum levels of \Red{ultracold neutrons}}

\author[a]{S. Bae\ss{}ler},
\author[b]{V. V. Nesvizhevsky},
\author[c]{G. Pignol},
\author[c]{K. V. Protasov},
\author[d]{A. Yu. Voronin}
\address[a]{University of Virginia, Charlottesville, VA 22904, U.S.A.}
\address[b]{Institut Laue Langevin, 6, rue Jules Horowitz, 38042 Grenoble, France}
\address[c]{Laboratoire de Physique Subatomique et de Cosmologie, 53, avenue des Martyrs, 38026 Grenoble, France}
\address[d]{Lebedev Physical Institute, Moscow, Russia}

\begin{abstract}
In this paper, we discuss a possibility to improve constraints on spin-dependent short-range 
interactions in the range of 1 -- 200\Unit{$\mu$m} significantly. For such interactions,  
our constraints are without competition at the moment. They were obtained through the observation of gravitationally bound states of ultracold neutrons. We are going to improve these constraints by about three orders of magnitude in a dedicated experiment with polarized neutrons using the next-generation spectrometer GRANIT.
\end{abstract}

\end{frontmatter}

\section{Introduction}

We propose an experiment to search for a new spin-dependent and short-range interaction. Such an interaction could be caused by new, light, pseudoscalar bosons such as the axion. The axion was originally proposed in \cite{Pec77a,Pec77b,Wein78,Wil78} as a solution to the strong CP problem, caused by the smallness of the neutron electric dipole moment. The axion would have profound consequences in cosmology and \Red{astrophysics} \cite{Kus08}, and \Red{the non-observation of these effects} limits the axion to have a mass in between 10\Unit{$\mu$eV} and 10\Unit{meV}. \Red{Modern earth-based searches for Axions are mostly based on} its two-photon-coupling \cite{Hag03}. In addition, an axion, or another new pseudoscalar boson should couple to fermions. An exchange of a virtual pseudoscalar boson gives rise to a macroscopic force between fermions. Of interest here is the search for a new CP violating interaction between a fermion and the spin of another fermion, given by (see \cite{Moo84}):
\begin{equation}
V_{\rm{SP}} (r) = \frac{g_{\rm S} g_{\rm P}}{8\pi}\frac{\left( {\hbar c} \right)^2 }{m_{\rm n} c^2}\left( {\Red{\mathbf\sigma _{\rm n}} \cdot \Red{\hat{\mathbf r}}} \right)\left[ {\frac{1}{{r\lambda }} + \frac{1}{{r^2 }}} \right]\eH{ -r / \lambda}
\label{eq:SPPotential}
\end{equation}
Here, $g_{\rm S}$, and $g_{\rm P}$ are the scalar and pseudoscalar couplings of the fermions to the exchange boson, where $g_{\rm P}$ acts on the polarized fermion. The range of this potential is determined by $\lambda = \hbar c/m$, where $m$ is the mass of the exchange boson. We note that more general forms of spin-dependent short-range interaction due to the exchange of new bosons are possible, as discussed in \cite{Dob06}, but are beyond the scope of our present discussion.

In a previous experiment, in which gravitationally bound states of ultra-cold neutrons were observed for the first time \cite{Nes02, Nes03, Nes05}, the height of the lowest bound states were visualized. Here, ultra-cold neutrons enter a slit between a flat horizontal mirror and an absorber/scatterer. At the exit, the transmission through the slit was measured as a function of the slit height $\Delta h$. The potential seen by a neutron above the mirror is the gravitational potential, which increases linearly with the height coordinate $z$. The height of the wave functions of the neutron states increases with the state number $k$. Neutrons whose wave functions are not vanishing at the absorber/scatterer were removed from the through-going neutron beam. Only the neutrons whose wave function are low enough could be detected at the end of the slit. An additional potential will change the shape of the wave function, and consequently of the neutron count rate at a given height of the slit. In principle, the presence of the absorber affects the shape of the wave function, too; but all models for the rough absorbers used in these experiments \cite{Nes05, Vor06, Wes06, Adh07} agree in that these changes are small (but not negligible). As the experimental results supported the quantum mechanical expectation without additional forces, the agreement can be turned into constraints onto additional forces. This has been done for spin-independent forces with a range of nanometers \cite{Nes04} or micrometers \cite{Wes07}. There is also the possibility to set a limit on the spin-dependent short-range forces of interest here, even if all particles are unpolarized. Adding a potential as in eq. \ref{eq:SPPotential} would cause the wave function of the ultra-cold neutrons to split into two spatially distinct components for the two possible orientations of the neutron spin. This is similar to a Stern-Gerlach experiment: One of these components would be slightly higher, and the other slightly lower than as in the case without an additional potential. The dependence of the neutron transmission versus the absorber height would look different and start at lower slit heights if a spin-dependent potential above our sensitivity limit would be present. This analysis is discussed in \cite{Bae07}.  

In this article, we will propose a dedicated experiment, in which we will use polarized neutrons to turn the absolute measurement of the transmission vs. slit height dependence into a relative measurement. The existence of an extra spin-dependent short range interaction would show up as a dependence of the neutron transmission through a slit with a given height $\Delta h$ on the polarization state of the neutron.

\section{The effect of an additional potential on the neutron transmission}

\begin{figure}[h]
  \begin{center}
    \includegraphics{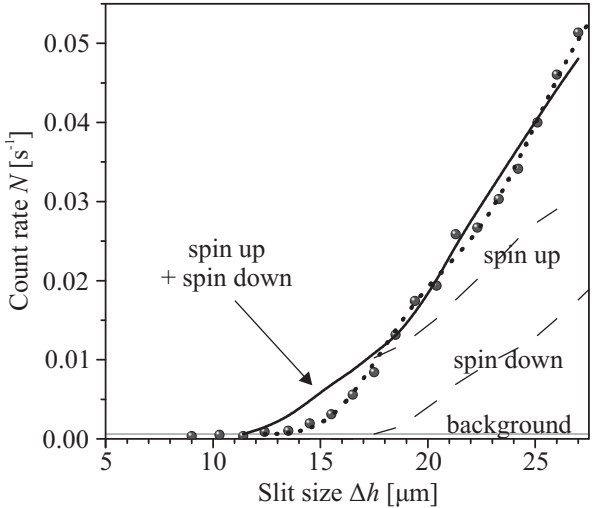} 
  \end{center}
\caption{Count rate through slit vs. slit height. Besides the data points, model functions without additional potentials (dots), with a new spin dependent potential for both spin states (dashed lines) and the sum of the last two (solid line) are shown.}
\label{fig:OldData}
\end{figure}

In the previous experiment, we could describe the measured neutron transmission through the slit between bottom mirror and absorber in the tunneling model. The data and different model functions are shown in fig. \ref{fig:OldData}.
The lifetime of a neutron in a given energy state $k$ is called $\tau_{k, \rm absorption}$ and has to be calculated using a model of the absorber. In the tunneling model, the neutron transmission for an ultra-cold neutron is given by
\begin{align}
T(\Delta h,k) & = \beta_k\eH{-\frac{\tau_{\rm passage}}{\tau_{k, \rm absorption}}} \label{eq:TunnelTransmission}
\\
& = \beta_k\left\{ 
  \begin{array}{ll}
    \eH{-\frac{L}{v_{\rm hor}}\cdot\alpha\cdot \exp\Klc{ -\frac{4}{3}\Kla{\frac{\Delta h - z_k}{l_0}}^{3/2}}}& \textrm{; }\Delta h > z_k \\
	\eH{-\frac{L}{v_{\rm hor}}\cdot\alpha\cdot 1} & \textrm{; otherwise}
  \end{array} \right. \nonumber
\end{align}
Here, $l_0$ is the characteristic length scale, which is
\begin{equation}
l_0  = \sqrt[3]{{\frac{{\hbar ^2 }}{{2m_{\rm n}^2 g}}}} = 5.87\Unit{$\mu$m}
\end{equation}
The $z_k$ are the classical turning points, and we have $z_1=2.34 l_0$, $z_2 = 4.09 l_0$, $\dots$ for the wave function unperturbed by the absorber. The factor $\beta_k$ allows for a suppression of individual states ($\beta_1 \sim 0.7$ is found from a fit to the data, all other $\beta_k$ are set to unity), and $\alpha$ parametrizes the efficiency of the absorber and the frequency with which the ultra-cold neutron attempts to cross the gravitational barrier. The passage time $\tau_{\rm passage}$ is given by the length of the absorber $L$ and the average horizontal velocity $v_{\rm hor}$ of the neutrons. The total transmission is then obtained by a sum over the transmission of all neutron states $k$ \cite{Wes02, Vor06}, truncated at some state number high enough that the transmission in this state can be neglected. The tunneling model is further described in \cite{Nes05}. 

\begin{figure*}
  \begin{center}
    \includegraphics{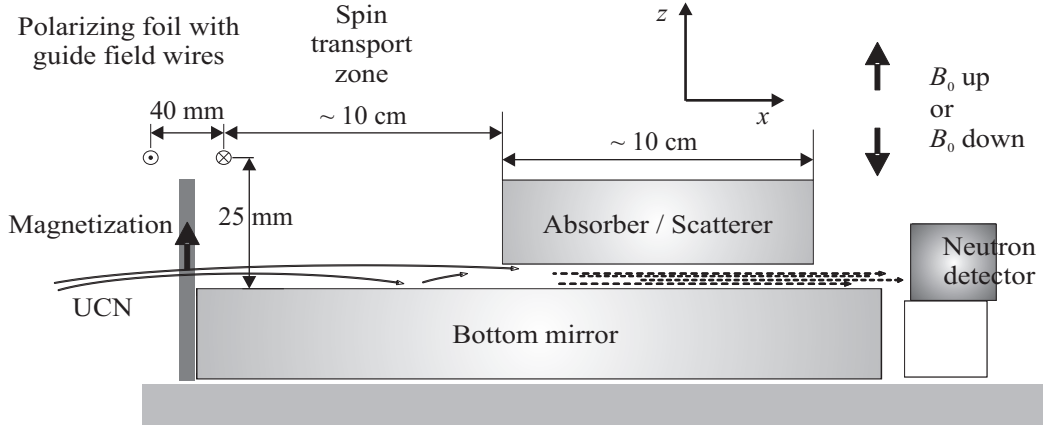} 
  \end{center}
\caption{Setup of the proposed experiment.}
\label{fig:ExperimentSetup}
\end{figure*}

The additional potential acting on the neutron can be gained by integrating eq. \ref{eq:SPPotential} over the mirror:
\begin{align}
V_{{\rm{SP}}} \Kla{z,\lambda ,\Delta h} = &+ g_{\rm S}^{\rm N} g_{\rm P}^{\rm n} \frac{\Kla{\hbar c}^2 \rho _{\rm m} \lambda }{8m_{\rm{n}}^2 c^2} \eH{ - \frac{z}{\lambda}}\underbrace {\Kla{\Red{\mathbf\sigma _{\rm n}} \cdot \Red{\hat{\mathbf z}}}}_{ \pm 1} \nonumber\\
&-g_{\rm S}^{\rm N} g_{\rm P}^{\rm n} \frac{\Kla{\hbar c}^2 \rho _{\rm m} \lambda }{8m_{\rm{n}}^2 c^2}\eH{ - \frac{\Delta h - z}{\lambda}}\underbrace {\Kla{\Red{\mathbf\sigma _{\rm n}} \cdot \Red{\hat{\mathbf z}}}}_{ \pm 1}
\label{eq:MAPotential}
\end{align}
The first term is caused by the mirror, and the second by the absorber. The density of mirror and absorber is given by the glass substrate, which might be quartz ($\rho _{\rm m} \sim 2.5$\Unit{g/cm$^3$}). As the coupling constants depend on the particle which is coupling, we denote with $g_{\rm S}^{\rm N}$ the scalar coupling to a nucleon of mirror or absorber and with $g_{\rm P}^{\rm n}$ the pseudoscalar coupling to the neutron. For sufficiently high $\lambda$, we find that the potential is linear in $z$ and we can calculate the shift of the classical turning points by a redefinition of gravitational acceleration $g$ (see the discussion in \cite{Bae07}). For arbitrary $\lambda$, we can use the WKB approximation\footnote{Our problem is discussed in \cite{Sak94,Wes02}, note the unusual constant $1/4$ which is due to the potential wall as the lower boundary condition.}: We calculate the energies $E_k^\pm$ and modified turning points $z_k^\pm$ through
\begin{gather}
\frac{{\sqrt {2m} }}{\hbar }\int\limits_0^{z_k^\pm}
{\sqrt {E_k^\pm \left( {g_{\rm{S}} ^{\rm{N}} g_{\rm{P}} ^{\rm{n}} ,\lambda } \right) - mgz - V_{{\rm{SP}}} \left( {z,\lambda ,\Delta h} \right)} } dz \nonumber \\
 = \pi \Kla{k - \frac{1}{4}}
\label{eq:BohrSommerfeld}
\end{gather}
The energy $E_k^\pm$ and the position of the classical turning point $z_k^\pm$ of each state $k$ depends on the orientation of the neutron spin. We end up with the spin-dependent transmission of the slit to be
\begin{equation}
N^\pm(\Delta h)=N_0 \sum_k T^\pm(\Delta h,k)
\label{eq:TotalCountRate}
\end{equation}
Here, $N_0$ is an overall normalization factor. $T^\pm(\Delta h,k)$ differs from $T(\Delta h,k)$ by the use of the shifted turning points $z_k^\pm$. \Red{A variant of this method was used in \cite{Wes07}.}

\section{The experimental setup}

\begin{figure}
  \begin{center}
    \includegraphics{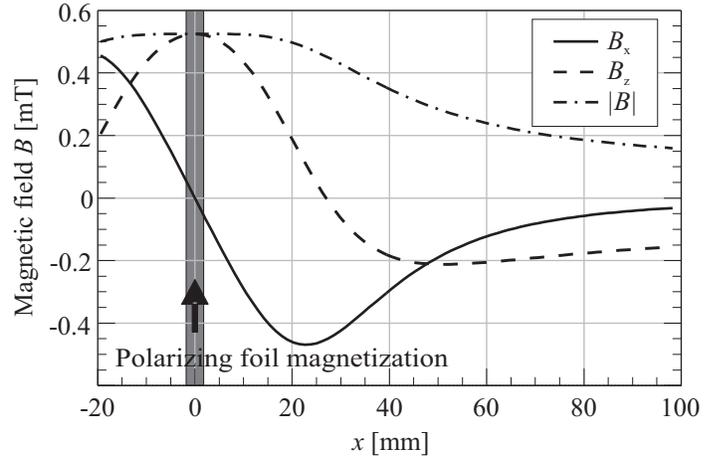} 
  \end{center}
\caption{Superposition of magnetic field of transport and guiding field.}
\label{fig:MagFieldMinusZ}
\end{figure}

In the new experiment, we will use a variant of the GRANIT experiment \cite{Nes06, Pign07, Pign08} to measure the transmission of polarized neutrons through the slit between a bottom mirror and an absorber. We will use the new ultracold neutron source \cite{SWell08} which is dedicated to the GRANIT experiment to make use of the higher statistics. We will have to polarize the ultracold neutrons (UCN) and maintain the polarization with a magnetic holding field. Therefore, the UCN pass a polarizing foil at the entrance of the spectrometer. The experiment is embedded in a homogeneous magnetic holding field $B_0$, whose size and magnitude can be changed for systematic reasons, as discussed below. The holding field is responsible for maintaining the spin state of the neutron. The setup is shown in fig. \ref{fig:ExperimentSetup}:

The polarizing foil is a silicon wafer, 300\Unit{$\mu$m} thick, with a ferromagnetic coating. After the polarizing foil, the neutrons are guided by a magnetic field behind the polarizer. This magnetic guiding field will keep the neutrons polarized along the magnetic field lines. It can be set in different ways:
\begin{itemize}
\item Polarization in {\bf +z} direction: Our guide field coils (not shown in fig. \ref{fig:ExperimentSetup}) produce a homogeneous magnetic field in +z direction with a size of $B_0 = 0.1$\Unit{mT}. The wires for the spin transport are not used.
\item Polarization in {\bf -z} direction: The wires for the spin transport are used, a current of $I=77$\Unit{A} runs through them. The magnetic field in this configuration is shown in fig. \ref{fig:MagFieldMinusZ}. At the end of the spin transport zone, the neutron spin points along the magnetic field lines, which is in -z direction. The neutron which traverses the polarizing foil sees a magnetic field which changes slowly enough that the spin stays aligned along the magnetic field lines.
\end{itemize}

After the neutron spin state is selected, the neutrons stay in the magnetic holding field $B_0$ while passing through the slit between the bottom mirror and an absorber. The transmission of the slit in the different spin states is given by eq. \ref{eq:TotalCountRate}. The turning points $z_k^\pm$ depend on the spin state, and as a result the count rates $\Np(\Delta h)$ and $\Nm(\Delta h)$ of the UCN behind the slit depend on the direction of the neutron spin if an additional potential as in eq. \ref{eq:SPPotential} would be present.

A special extra-low background gaseous He-3 neutron counter has been developed and tested for this experiment. Neutron velocity spectra will be measured using position-sensitive UCN detectors analogous to those presented in \cite{Nes00}.

\section{Discussion of expected uncertainties} 
\begin{figure}
  \begin{center}
    \includegraphics{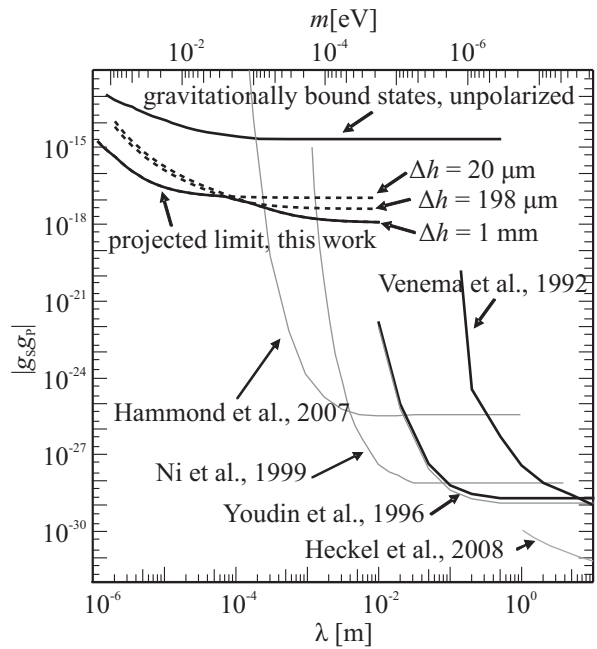} 
  \end{center}
\caption{Constraints from different experiments on a new spin-dependent short-range interaction. Note that the definition of the coupling constants is different for different authors by a factor of \Red{$\hbar c$}. The \Red{thick} black lines are constraints for an interaction coupling to a neutron spin, and the thin gray lines are for interactions coupling to an electron spin. \Red{The projected limits are given as the envelope of measurements at three different absorber heights $\Delta h$.}}
\label{fig:Limits}
\end{figure}

We measure the count rate asymmetry $\epsilon(\Delta h)$ for different directions of the neutron spin, defined as
\begin{equation}
\epsilon(\Delta h)=\frac{\Np(\Delta h)-\Nm(\Delta h)}{\Np(\Delta h)+\Nm(\Delta h)}
\end{equation}
Compared to our previous experiment, which was limited by systematics, this asymmetry is more sensitive to turning point shifts due to the relative nature of the measurement. Therefore, we expect to gain a factor of 30 in the sensitivity compared to our previous analysis. In addition, we gain in statistics a factor of 3 due to the larger mirrors used in GRANIT, and a factor of 15 due to a more intense source and maybe a longer measurement time. \Red{In fig. \ref{fig:Limits}, the projected sensitivity limit is shown for three different settings of the absorber heights $\Delta h$. If we wanted to achieve the optimum sensitivity for interaction ranges $\lambda > 1\Unit{mm}$, we would do additional measurements with $\Delta h \sim \lambda$. However, we would not be competitive. Fig. \ref{fig:Limits} shows also limits from other experiments. } Only in \cite{Ven92, You96}, a limit could be extracted for a pseudoscalar coupling to a neutron. \Red{As a comparison, constraints from \cite{You96},} and from others \cite{Ni99, Hamm07, Heck08}, for a pseudoscalar coupling to an electron are also shown. 

The most important systematic uncertainties are connected with the magnetic field. In the linear approximation, the proposed additional interaction from eq. \ref{eq:MAPotential} with a magnitude at the limit of our sensitivity looks like an apparent magnetic field gradient with a size of the order of 1\Unit{$\mu$T/cm} (We are not sensitive to a constant magnetic field). This is a magnetic field gradient which is easily avoided if its origin is an imperfection of the magnetic holding field $B_0$. Ferromagnetic impurities in bottom mirror and/or absorber could in principle mimic such a field at least in average, but their main effect would be some kind of magnetic roughness which would cause a loss of all UCN, as the roughness would cause transitions between different energy states. In addition, such inhomogeneities would hardly be homogeneous over the width of the slit. In addition to these indication, we are planning to look for ferromagnetic impurities with other methods, i.e. SQUIDs (Superconducting Quantum Interference Devices).
The effect of para- or diamagnetism in mirror or absorber is expected to be small. To the first order, it will add a homogeneous field to the magnetic holding field $B_0$ which does not influence the result.
False effects due to the magnetic interaction with mirror and absorber can be studied in measurements with different sizes or orientations of the magnetic holding field or the magnetization of the polarizer. In addition, we can repeat our experiment for different slit sizes $\Delta h$, for one to pinpoint the range of an additional interaction if we find it, and to discriminate against possible magnetic false effects.

We can prove the adiabaticity of spin transport across the installation by adding an analyzer to the setup.

There is a small Stern-Gerlach effect in the spin transport zone, which might shift the UCN
by about a micrometer. This is a potentially dangerous systematic, as it can couple to a potential position-dependence of the efficiency of the polarizer.
Again, a change of the size of the magnetic holding field $B_0$ will affect the Stern-Gerlach effect, while it will not influence an effect due to an additional short-range interaction.

\section{Conclusions}

We interpreted the results of the experiment, in which gravitationally bound neutron states were discovered, as constraints on extra spin-dependent short-range interactions which could be caused by axion-like particles. These constraints can be significantly improved if polarized neutrons are used. Such an experiment will be performed in the setup phase of the new GRANIT spectrometer. This type of experiment is uniquely suited for \Red{a search for new interactions} with a range of $1-200$\Unit{$\mu$m}, as this is the characteristic length scale of the lowest bound quantum states. 



\end{document}